\documentclass[preprint, aip, rse]{revtex4-1}

\usepackage{CJK}

\usepackage{graphicx}
\usepackage{mhchem}

\begin{document}

\begin{CJK*}{GB}{}

\author{A. Carati}
\affiliation{Universit\`a degli Studi di Milano, Dipartimento di
  Matematica, via Saldini 50, 20133 Milano (Italy).}
\author{M. Marino} 
\affiliation{Universit\`a degli Studi
  di Milano-Bicocca, Dipartimento di Scienze della Salute, via Cadore
  48, Monza (Italy).}
\author{D. Brogioli} 
\affiliation{Universit\`a degli Studi
  di Milano-Bicocca, Dipartimento di Scienze della Salute, via Cadore
  48, Monza (Italy).}
\email{dbrogioli@gmail.com}

\title{Theoretical thermodynamic analysis of a closed-cycle process
  for the conversion of heat into electrical energy by means of a
  distiller and an electrochemical cell.}

\begin{abstract}
We analyse a device aimed at the conversion of heat into electrical
energy, based on a closed cycle in which a distiller generates two
solutions at different concentrations, and an electrochemical cell
consumes the concentration difference, converting it into electrical
current.  We first study an ideal model of such a process. We show
that, if the device works at a single fixed pressure (i.e. with a
``single effect''), then the efficiency of the conversion of heat into
electrical power can approach the efficiency of a
reversible Carnot engine operating between the boiling temperature of
the concentrated solution and that of the pure solvent.  When two heat
reservoirs with a higher temperature difference are available, the
overall efficiency can be incremented by employing an arrangement of
multiple cells working at different pressures (``multiple effects'').
We find that a given efficiency can be achieved with a reduced number
of effects by using solutions with a high boiling point elevation.
\end{abstract}

\maketitle

\end{CJK*}

\section{Introduction}
\label{sect:intro}

The free energy contained in solutions with different concentrations
can be tapped and converted into electrical energy, e.g. by means of a
concentration difference electrochemical cell~\cite{cole1964}. The use
of naturally occurring solutions with different concentrations has
been proposed for the production of completely clean and renewable
energy~\cite{pattle1954, norman1974, logan2012}. Such solutions
include river and sea waters, and brines from salt lakes (e.g. Dead
Sea)~\cite{loeb1998}, from coal-mines (produced by dissolving
geological deposits)~\cite{turek2008}, and from
salterns~\cite{cipollina2012}.

Various techniques have been proposed for the conversion of the
concentration differences into current. In the so-called
pressure-retarded osmosis~\cite{levenspiel1974, loeb1975, chung2012}
technique, a semi-permeable membrane is interposed between the
solutions with different concentrations, generating an osmotic water
flow, which is sent to a turbine. In reverse electrodialysis
(RED)~\cite{weinstein1976, post2008, post2007} the two feed solutions
are sent to a stack of ion-exchange membranes; the ion diffusion
across the membranes constitutes a current that can be extracted from
the cell. Other techniques are based on electrochemical concentration
cells~\cite{cole1964, clampitt1976} or on the concept of the
accumulator-mediated mixing, i.e. capacitive mixing
(CAPMIX)~\cite{brogioli2009, sales2010, brogioli2011, burheim2011,
  bijmans2012, rica2013, brogioli2013} and ``mixing entropy
battery''~\cite{lamantia2011, jia2013}.  In this paper, we will call
``salinity gradient power'' (SGP) cell any of the cells working with
one of the above-mentioned techniques.

\begin{figure}
\includegraphics{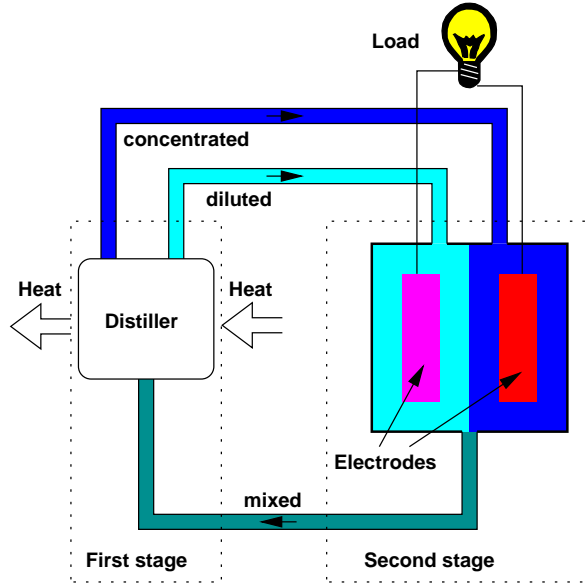}
\caption{Scheme of the closed-cycle device for the conversion of heat
  into electrical current (heat-to-current converter, HTCC). The
  solutions with different concentrations enter into the
  SGP cell, where are used for producing electrical
  current. The outgoing flows, that are completely mixed, are sent
  to the distiller, where their concentration difference is restored.
\label{fig:htcc}}
\end{figure}

The idea that is investigated in the present paper is to apply these
techniques to solutions with different concentrations that are
produced by means of distillation, as shown in Fig.~\ref{fig:htcc}.
The solutions with different concentrations enter into the SGP cell,
and exit with a reduced concentration difference. The solutions are
then sent to the distiller, where their concentration difference is
restored, and are sent again to the cell. The process is a closed
cycle, in which the solutions are regenerated, and the whole device
can be considered as a heat-to-current converter (HTCC), suited for
exploiting low-temperature heat sources. This idea has been already
suggested in \cite{lamantia2011} and a similar HTCC, involving
distillation of a ammonia/\ce{CO2} solution and a pressure-retarded
osmosis device feeding a turbine, has been already
proposed\cite{mcginnis2007}.

Such a conversion is traditionally obtained by means of heat engines
(see for example \cite{tlili2008}), e.g. Stirling motors or organic
Rankine cycle devices.  However, the quest for alternative methods,
less relying on moving parts requiring maintenance, is an important
research field. Alternative solutions for the heat-to-current
conversion include the thermoelectric elements based on the Seebeck
effect~\cite{snyder2008, weidenkaff2008, frassie2013}: although very
compact, they have a lower efficiency than the heat engines, and are
quite expensive. The HTCC we propose has advantages over the known
solutions: thanks to the limited presence of moving parts, it is
suited for domestic and small-scale applications.

The required heat can be obtained by a renewable source, e.g. by means
of a solar concentrator. Low levels of solar concentration can be
achieved~\cite{kaushika2000} by means of the concentrators working
with the so-called non-imaging optics~\cite{muschaweck2000,
  spirkl1998}. Such solar concentrators, not requiring tracking, are
suited for domestic applications. By using selective absorbers, the
heat collectors can work with good efficiency at moderately high
temperature, for example 50\% at 180$^{\circ}$C. The HTCC we propose
will allow the collected to be converted heat into electrical
current. Other applications include co-generation~\cite{omer2000,
  rosato2013, pan2013} and recovery of waste heat from industrial
processes. Our technique thus contributes to the renewable and clean
energy production.

The choice of the solution to be used in the closed-cycle HTCC that we
propose depends on the technology of both the SGP cell and the
distiller. The solvent is not limited to water, and can be an organic
liquid. Particular solutes have been proposed, e.g.\ lithium
salts~\cite{lamantia2011}.

The present work aims at finding a fundamental thermodynamic
limitation for the operation of our HTCC, that allows one to select
suitable solutions. We do not study a particular scheme for the SGP
cell, nor for the distiller. We assume that the SGP cell will be
developed by using one of the above-described known technologies. All
the described SGP technologies allow one to reach, at least in
principle, a good efficiency in the conversion of the free energy of
the feed solutions into electrical energy.  The goal of the present
paper is to show that a single property, namely the
boiling point, represents a clue for selecting a solution, without
having to carry out the design of the distiller and of the SGP
cell. This allows one to decouple the design of the distiller
from that of the SGP cell, so that the two stages of the HTCC 
can be studied independently of one another.

We first focus on a HTCC containing a single effect, i.e. working at a
single pressure. As an example, we present a practical scheme of a
single-effect device, working in a continuous regime. In order to give
a rough evaluation of the efficiency, we study an idealization of such
device, constituted by a cyclic process. We find a quite surprising
result: the efficiency of the HTCC increases when solutions with a
high boiling point elevation are used, contrary to the usual rule that
the efficiency of the distillation is reduced by the boiling point
elevation.  Multiple effects are often used in order to increase the
efficiency of distillation: we analyze the application of multiple
effects to our case, and we find that, given a temperature difference,
the same conversion efficiency can be obtained with less effects when
solutions with a higher boiling point elevation are used. Finally, we
show that the thermal properties, relevant for the design of the
distiller, also allow one to predict the electrical potential of the
electrochemical SGP cell.

\section{Example of realization of the single-effect process}
\label{sect:one:effect:practical}

\begin{figure}
\includegraphics{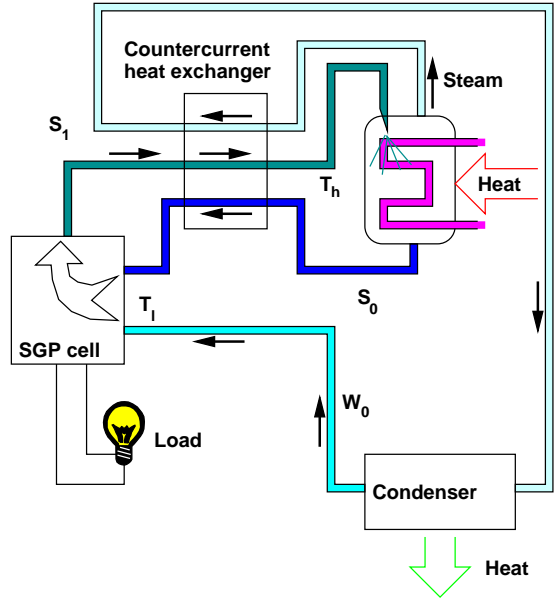}
\caption{Practical realization of the HTCC with a continuous process.
  \label{fig:one:effect:practical}}
\end{figure}

In Fig.~\ref{fig:one:effect:practical}, we present a practical scheme of
the HTCC, with a single evaporator (``single effect''). At the working
pressure, that is constant in all the parts of the device, the
concentrated solution and the pure solvent have respectively boiling
temperatures $T_h$ and $T_l$, with $T_h>T_l$~\cite{hala1967}. 

The SGP cell is fed with the concentrated solution $S_0$ and the
distilled solvent $W_0$.  The SGP cell generates a voltage, that is used
for powering a load. The flow of current through the SGP cell induces
the migration of the solute from the more concentrated to the less
concentrated solution. The SGP cell is designed so that it consumes
nearly completely the available concentration difference, while
working in a continuous regime (see \cite{kim2013} for an example
applied to RED technology), and gives as an output a completely 
mixed solution $S_1$.

The SGP cell works at temperature slightly lower 
than $T_l$, in order to avoid that the
pure solvent $W_0$ boils. The output solution $S_1$, at temperature
$T_l$, is first sent to the countercurrent heat exchanger, where it
receives heat; then it is sent to the evaporator, where additional
heat is supplied, until the temperature $T_h$ is reached and part of
the solvent evaporates. The concentrated solution $S_0$, at
temperature $T_h$, is sent back through the countercurrent heat
exchanger. If necessary, it is still further cooled, until the temperature
$T_l$ is reached, and finally comes back into the SGP cell.

The steam is condensed at temperature $T_l$, and the obtained
distilled solvent is sent to the SGP cell. In principle, it is also
possible to send also the steam through the countercurrent heat
exchanger.

We assume that the flow of distilled solvent $W_0$, that is sent to the
cell, is much smaller than the flow of the incoming solution $S_0$, so
that the concentration of the outgoing solution $S_1$ is close to the
concentration of $S_0$, and their boiling point is nearly the same.

\section{Thermodynamical efficiency of a single-effect cycle}

\begin{figure}
\includegraphics{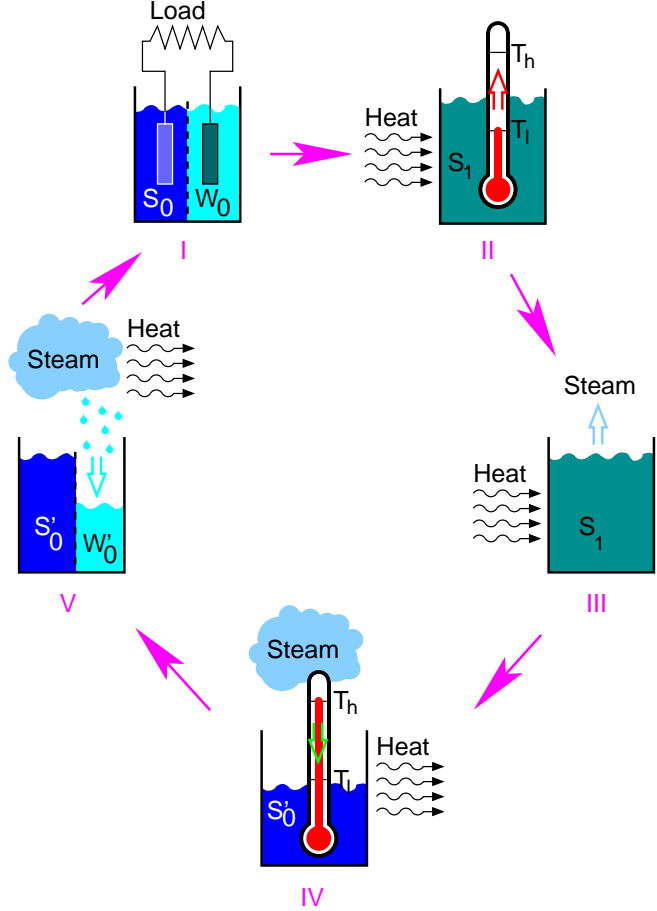}
\caption{Ideal implementation of the HTCC by means of a cycle.
\label{fig:one:effect:theo}
}
\end{figure}

In the present section, we evaluate the thermodynamic efficiency
$\eta$ of the HTCC, defined as the ratio between the produced
electrical energy and the total amount of heat absorbed from the heat
source during a cycle.

An upper bound for the efficiency of the above-described device is given by
the efficiency of the Carnot cycle~\cite{zemansky1957} between the
temperatures $T_h$ and $T_l$:
\begin{equation}
\eta < \frac {T_h- T_l}{T_h} = \eta_C .
\end{equation}
We will now show that the actual efficiency $\eta$ of the device
can indeed be very close to this upper
bound $\eta_C$. We consider the cycle sketched in Fig.~\ref{fig:one:effect:theo},
that represents an idealization of the operation of the HTCC
described in the previous section. Although such a cycle appears to be too
much idealized and impractical, its efficiency can be seen as a quick and
rough estimate of the target we aim at when designing a real HTCC.

The phases of the cycle are the following.
\renewcommand{\theenumi}{\Roman{enumi}}
\begin{enumerate}
\item The cycle starts with an amount of pure solvent $W_0$ in the first
  compartment of the SGP cell, and an amount of solution $S_0$ at
  concentration $c_0$ in the second one. We exploit the
  difference of potential to make an electric current flow through an
  external load. The flow of current induces a transfer of solute from the more
  concentrated solution to the less concentrated one. The process
  continues until at the end the concentrations in the two
  compartments become equal to a common value $c_1<c_0$.
\item The mixed solution $S_1$ is heated from $T_l$ to $T_h$. We suppose
  that the mass of the pure solvent $W_0$ is much smaller than that of
  $S_0$ (this feature is not exhibited in Fig.~\ref{fig:one:effect:theo}),
  so that the difference between $c_0$ and $c_1$ is small, and the
  boiling temperature does not appreciably vary during the
  evaporation.
\item An amount of the vapor of the solvent, having the same mass as
  that of the pure solvent $W_0$ initially present in the first compartment,
  is separated from the rest of the solution. This rest, which will be
  called $S_0'$, has the same mass and concentration $c_0$ as those of
  the solution $S_0$ initially present in the second compartment.
\item Both the solution $S_0'$ and the solvent vapor are cooled down to
  the temperature $T_l$, at which the vapor is condensed.
\item The vapor is condensed back to liquid solvent, and both the
  solvent and the solution $S_0'$ are ready to be transferred again to
  the concentration cell, where the cycle restarts.
\end{enumerate}

We assume that the performance of the SGP cell is ideal, i.e. that the
total amount of produced electrical energy is equal to the difference
in Gibbs free energy between the initial state (with the pure solvent
$W_0$ at one side and the solution $S_0$ at the other) and the final
one (with the same concentration $c_1$ at the two sides).

The main heat exchanges take place during the state changes, i.e. the
in phases II and IV, at temperatures $T_h$ and $T_l$ respectively. If
we simply neglect the heat needed for changing the temperature of the
solutions and of the vapor (phases II and IV) and during the
isothermic operation of the SGP cell (phase I), we conclude that
the cycle is reversible, and thus $\eta=\eta_C$.

A more rigorous evaluation of the heat exchanged in phases II and IV
(which is performed in the Supporting Information
(SI)\cite{supplemental}, Appendix A), shows that, in typical
situations, $\eta$ is very close to $\eta_C$ ($\eta_C -\eta<0.3$\%).

In a practical device, for example the one described in the previous
section, the irreversibility (i.e. the loss of free energy) mainly
comes from the non-ideal heat exchange in the heat exchanger and from the
incomplete extraction of free energy in the SGP cell. 

However, some of the features of the ideal cycle actually hold also in
a practical realization. In particular, we can notice that the
efficiency depends on the two boiling temperatures of the solutions,
and not on the actual temperatures of the heat reservoirs. In this
sense, if we want to increase the efficiency of the HTCC, we must
choose a solution with a boiling point elevation as high as possible,
compatibly with the temperatures of the available heat sources. This
concept is contrary to the widespread idea that the boiling point
elevation reduces the efficiency of distillation.

For water at atmospheric pressure one has $T_l=373$~K. Therefore, using a
solution having a boiling point elevation $\Delta T = T_h- T_l =$
40~K, we reach an efficiency of about 10\%, which is comparable to
that of photovoltaic devices.  Obviously, the HTCC can operate at a
pressure lower than the atmospheric one, thus reducing both $T_h$ and
$T_l$. It is worth noting that, thanks to the D\"uhring
rule~\cite{white1930}, the efficiency of the HTCC does not depend on
pressure, because both the boiling point temperatures change
proportionally.

\section{The HTCC with multiple effects}
\label{sect:two:effects}

Multiple effects are often used in order to increase the efficiency of
distillation. Basically, the idea is to use the heat that exits from
one of the stages (called effects) to feed the following one. In this
way, the same quantity of heat passes through the $n$ effects and is
thus used $n$ times.

\begin{figure}
\includegraphics{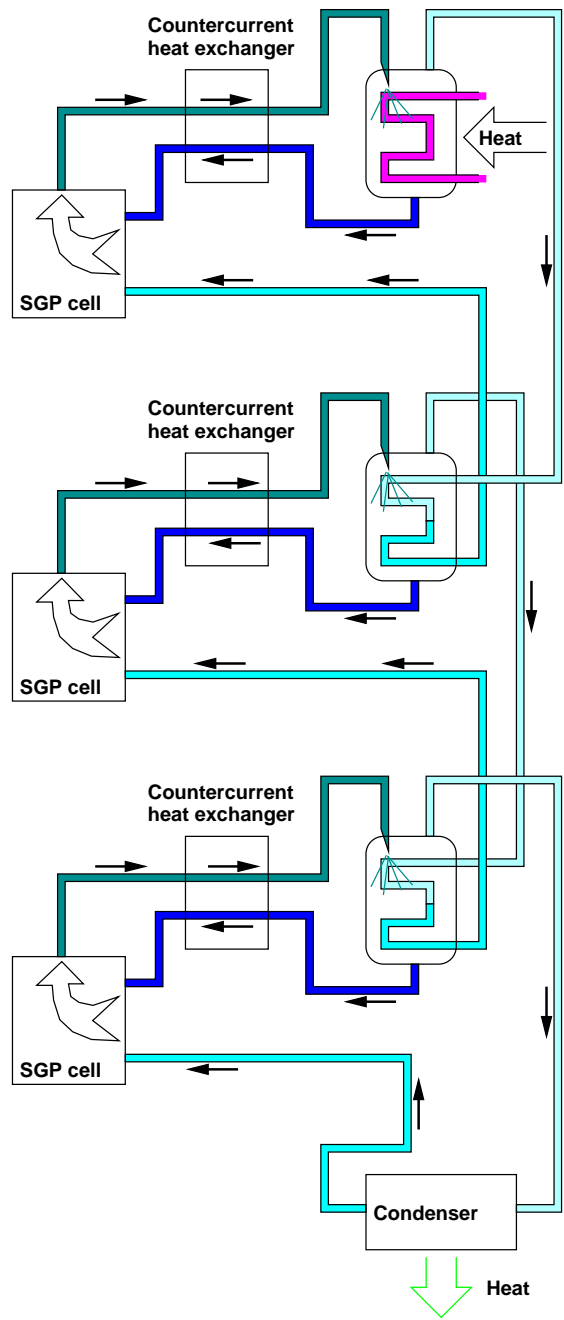}
\caption{Scheme of a practical HTCC with three effects.
\label{fig:multiple:effects}}
\end{figure}

In the present section we describe the application of the concept of
the multiple effects to the HTCC. A three-effect version of the HTCC is
shown in Fig.~\ref{fig:multiple:effects}. Each of the effects is similar to the
above-described practical HTCC.

As already stated, that the main part of the heat absorbed by one of the
effects is represented by the evaporation heat of the solution at the
temperature $T_h$, whereas the main release of heat at the temperature
$T_l$ corresponds to the condensation heat of the solvent vapor.  We
couple a first stage (effect) to a second one, similar to the first
one but working at a lower pressure, such that at this lower pressure
the boiling temperature of the solution is equal to $T_l$. Then the
second stage will produce additional electrical energy by using the
heat released by the condensation of the solvent vapor that goes out of
the first one. Generalizing this idea, we can realize $n$ similar
devices, all working with solutions of the same concentration $c$, but
at pressures $P_1> P_2 > \cdots > P_n$ such that, denoting by
$T_h(P)$ and $T_l(P)$ respectively the boiling temperatures of the
solution and of pure solvent at pressure $P$, one has
$T_h(P_2)=T_l(P_1):=T_1$, $T_h(P_3)= T_l(P_2):= T_2$ and so on. The
first device absorbs the heat $Q_0$ at the temperature $T_h(P_1):=
T_0$, and delivers the heat $Q_1$ to the second device at the
temperature $T_1$. Heat exchanges of the same type occur between all
of the subsequent devices of the chain, until the last one releases the
heat $Q_n$ to the external environment at the temperature $T_n$, which
may be equal to room temperature if $P_n$ is sufficiently low.

By using the same approximations used above, one can assume that the
total system including all the $n$ devices mainly absorbs from the
external source only the heat $Q_0$ at the temperature $T_0$, and
releases to the environment only the heat $Q_n$ at the temperature
$T_n$. We know that each effect can work close to reversibility, and
thus the overall efficiency $\bar \eta$ of the whole multiple-effect
HTCC is given by
\begin{equation} \label{carnotm}
\bar\eta \approx \frac {T_0- T_n}{T_0}
\end{equation}
Thus we see that the multiple-effect scheme is useful when we want to
exploit two heat reservoirs with a temperature difference $T_0-T_n$
that is larger than the boiling point elevation $T_h-T_l$ of the used
solution.

On the other hand, in each effect the temperature difference
$T_{n-1}-T_n$ mirrors the boiling point elevation at the working
pressure $P_n$. From this point of view, the use of solutions with a
high boiling point elevation allows us to reduce the total number of
effects needed to fully exploit a given temperature difference
$T_0-T_n$.

\section{The potential of the concentration cell}
\label{sect:potential}

The result that the efficiency increases with the boiling point
elevation comes as a surprise because the presence of a high boiling point
elevation is usually connected with a lower production of distilled
products. This is due to the peculiarity of the application we
consider. Indeed, a higher
boiling point elevation results in a higher free energy content per
unit mass of produced solution, and thus to an increase of efficiency,
associated to a decrease of the total flow of liquids. In the present
section we discuss the relation between the thermal properties
(i.e. the boiling points and the vapor pressures of the solutions) and
the potential of the electrochemical cell, or, better, the chemical
potential of the solute.

With reference to the single-effect ideal cycle described above, let
us assume that the solution $S_0$ contains $M$ moles of solvent and $N$
moles of solute, 
$W_0$ contains $m$ moles of solvent,
and $S_1$ threfore contains $M+m$ moles of solvent and $N$ moles of
solute. 
The Gibbs free energies of the solutions $S_0$, $W_0$ and $S_1$
at temperature $T$ are  $G\left(M,N,T\right)$,
$G\left(m,0,T\right)$ and $G\left(M+m,N,T\right)$ respectively. 
The electrical work
$W$ produced by the SGP cell equals the total available Gibbs free
energy:
\begin{equation}
W = G\left(M,N,T\right) + G\left(m,0,T\right) - G\left(M+m,N,T\right) .
\end{equation}
If $m\ll M$, the above relation can be expressed in terms of the chemical 
potential of the solvent
$\mu_A\left(c,T\right)$ as
\begin{equation}
W = m\left[ \mu_A\left(0,T\right) - \mu_A\left(c,T\right) \right] .
\end{equation}
So, expressing $W$ through the definition $W=\eta Q$ of efficiency, where
$Q$ is the amount of heat absorbed from an external source, one gets
\begin{equation}
\mu_A\left(0,T\right) - \mu_A\left(c,T\right) = \frac{Q}{m} \eta .
\end{equation}
Since in our case $Q$ is roughly the heat needed for 
evaporating $m$ moles of the
solvent, we can write $Q=m\lambda$, where $\lambda$ is the latent 
heat of evaporation of the solvent. 
Finally, substituting the value of $\eta$ that was previously evaluated,
one obtains
\begin{equation}
\mu_A\left(0,T\right) - \mu_A\left(c,T\right) = \lambda \frac{T_h-T_l}{T_h} ,
\end{equation}
Thus we see that the molecules of the solvent feel a stronger chemical
potential gradient between the concentrated solution and the pure
solvent when the boiling point elevation is higher. It is worth noting
that the difference of the chemical potential of the solvent between
the solutions is proportional to the osmotic pressure, that is
exploited in the SGP devices based on osmosis.

Now, using the formula above, the boiling point elevation can be
related to the output voltage of an electrochemical SGP cell. Indeed,
we can evaluate the voltage by means of the Nernst equation:
\begin{equation}
\label{eq:nernst}
V= \frac{\mu_S(c_2,T)- \mu_S(c_1,T)}{zF} ,
\end{equation}
where $\mu_S$ is the chemical potential of the solute, $z$ is the
number of electrons exchanged during the redox reaction, and $c_1$ and
$c_2$ are the concentrations of the solutions in the SGP cell. In
turn, $\mu_S$ can be evaluated from $\mu_A$, by means of the 
relation
\begin{equation} 
\label{eq:colleg}
(1-c)\frac{\partial \mu_A}{\partial c} + c\frac{\partial \mu_S}
{\partial c} =0 .
\end{equation}
The derivation of this equation can be found in SI\cite{supplemental},
Appendix B, together with a more rigorous procedure for the
calculation of the chemical potential from the boiling point
elevation. The concepts introduced in this section are also developed
in SI\cite{supplemental}, Appendix C, to obtain an alternative
derivation of the efficiency for the ideal cycle.

\begin{figure}
\begin{tabular}{c}
a\\
\includegraphics{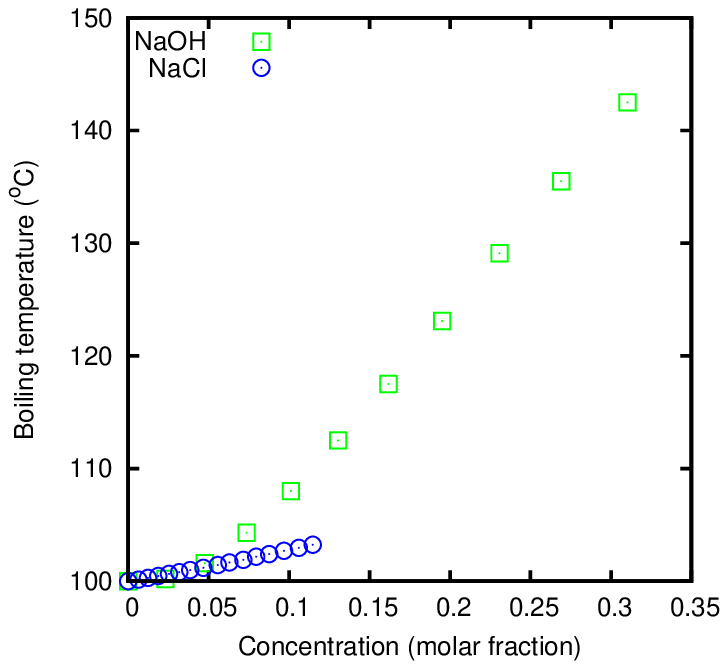}
\\
b\\
\includegraphics{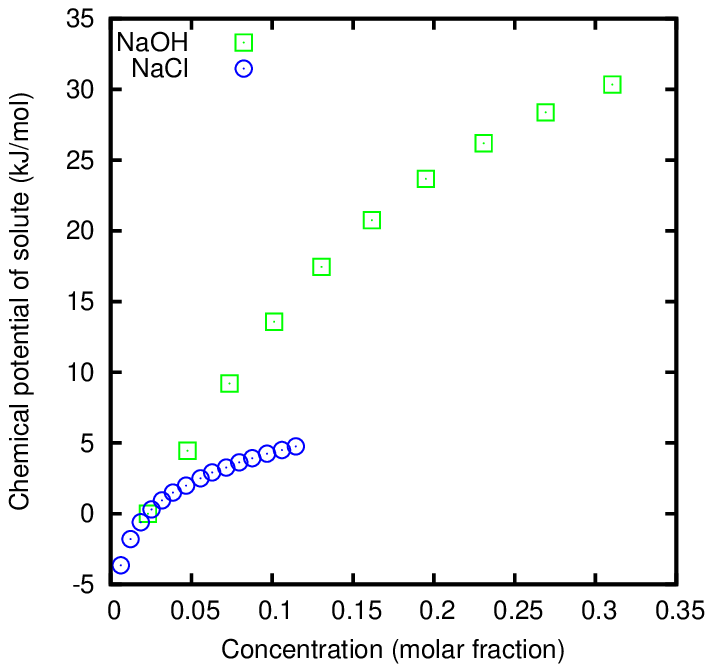}
\end{tabular}
\caption{Panel a: boiling temperature at atmospheric pressure of aqueous
solutions of \ce{NaOH} and \ce{NaCl}. The data for the \ce{NaCl} stop
at the molar fraction corresponding to the solubility of the salt.
Panel b:Chemical potentials of solute for aqueous solutions of
\ce{NaOH} and \ce{NaCl} at $T=100^\circ$C. The data for the \ce{NaCl} stop
at the molar fraction corresponding to the solubility of the salt.
} 
\label{fig:boiling:point}
\end{figure}

To give examples of application of this procedure, we show in
Fig.~\ref{fig:boiling:point}a the curves of the boiling temperatures at
atmospheric pressure of aqueous solutions of NaOH and NaCl, and in
Fig.~\ref{fig:boiling:point}b the resulting curves of the chemical
potential of the solute at the temperature $T=100^\circ$C. Both curves
have been arbitrarily set to 0 at the concentration $c=0.023$, which
corresponds to 5\% in weight for NaOH.  Note in fact that
Eq.~\ref{eq:colleg} allows us to evaluate the chemical potential only
up to an additional constant term, which is however irrelevant for the
calculation of the potential of the cell. One can notice how the
higher boiling point elevation of NaOH entails a steeper increase,
with respect to NaCl, of the chemical potential as a function of the
concentration.  By applying Eq.~\ref{eq:nernst} we obtain that a concentration
cell, having in a compartment a solution of NaOH at 5\% in weight, and
in the other one a solution of NaOH at 50\% in weight (which
corresponds to a molar fraction of 0.31 and a boiling point elevation
at atmospheric pressure of $42^\circ$C), provides in the ideal case a
potential of 0.31 V at $T=100^\circ$C.

\section*{Acknowledgements}
We thank Alessandro Atti, Pino Gherardi and Paolo Turroni
for the collaboration in the development of the concept of the
closed-cycle heat-to-current converter. We thank Luigi Galgani and
Maarten Biesheuvel for useful discussions and suggestions on the
manuscript.  The research leading to these results received funding
from the European Union Seventh Framework Programme (FP7/2007-2013)
under agreement no. 256868, CAPMIX project.

\end{document}